# Food Recommendation using Ontology and Heuristics

M. A. El-Dosuky[1], M. Z. Rashad[1], T. T. Hamza[1], A.H. EL-Bassiouny[2]

[1] Dep. of Computer Sciences, Faculty of Computers and Info. Mansoura University, Egypt,
[2] Dep. of Mathematics, Faculty of Sciences, Mansoura University, Egypt
[1]{mouh_sal_010, magdi_12003,Taher_Hamza}@mans.edu.eg, [2]el_bassiouny@mans.edu.eg

**Abstract.** Recommender systems are needed to find food items of one's interest. We review recommender systems and recommendation methods. We propose a food personalization framework based on adaptive hypermedia. We extend Hermes framework with food recommendation functionality. We combine TF-IDF term extraction method with cosine similarity measure. Healthy heuristics and standard food database are incorporated into the knowledgebase. Based on the performed evaluation, we conclude that semantic recommender systems in general outperform traditional recommenders systems with respect to accuracy, precision, and recall, and that the proposed recommender has a better F-measure than existing semantic recommenders.

## 1 Introduction

Recommender systems are needed to find food items of one's interest. Challenges in building nutrition recommender systems can be classified as those concerning the user, and those concerning the algorithms used [1]. Different models are proposed [2] to deal with the missing or incorrect data from food recording measurements. Other challenges have a trade-off between them such as the perfect databases size and the cold-start problem. The cold-start problem can be solved by using information about the user's previous meals to calculate similarity measures to recommend new recipes [3]. Challenges about user compliance can benefit from many suggested strategies[4]. Users need nutrition heuristics to help develop a bias toward eating healthfully [5].

Section 2 reviews the previous attempts in building food recommenders and recommendation approaches. Section 3 presents our solution and the evaluation of the proposed framework. We conclude in Section 4 with plans for future work.

## 2 Previous Work

First efforts of designing automated systems to plan a meal based on personal nutritional needs utilize case-based planning such as CHEF [6] and JULIA [7]. A recipe recommender system usually employ similarity measures to recommend recipes that are most similar to meals the user likes [3]. User ratings are core for the recommender system [8], taking into account heuristics indentified by health care providers [9]. A system that analyses shopping receipts and then recommends healthier food choices is proposed [10]. To calculate the nutritional content of meals, Smart Kitchen [11] is proposed. Computer vision can be applied to analyze pictures of meals to predict the nutritional content [12]. Other systems focus on analysing the written form of nutritional content ([13], [14]). Recent attempts try to improve recipe recommendations by understanding the user's tastes [15].

There are four types of recommender approaches: content-based, semantics-based, collaborative filtering, and hybrid [16], but we restrict our discussion to the first two only. Content-based recommenders make use of Term Frequency-Inverse Document Frequency (TF-IDF)[17] and cosine similarity to compare the similarity between documents. Semantics is concerned only with concepts, and employing approaches such as concept equivalence [18], binary cosine[18], Jaccard [19], and semantic relatedness [20]. Next section shows how these approaches can be implemented.

## 3   Proposed Framework

The proposed framework is shown in fig. 1.

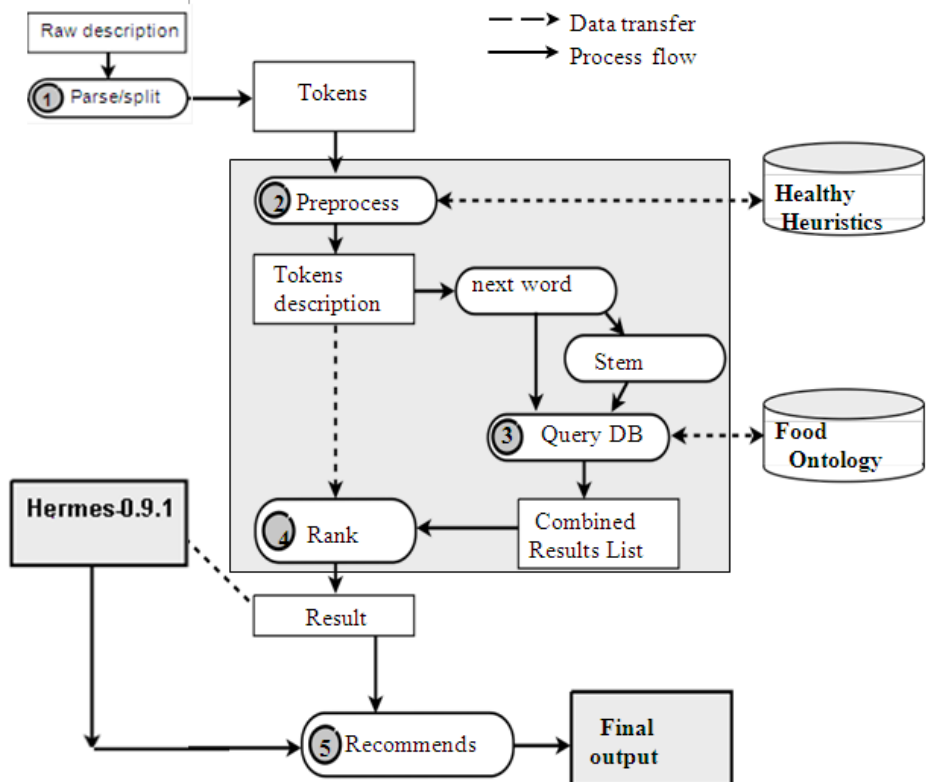

**Fig. 1**. The proposed framework

The first step is to take the raw description directly from the user or from his profile. Stop words are removed, followed by stemming words back to the root and removing punctuation and converting to lower case. To develop a bias toward healthful food, examined nutrition heuristics are collected [5]. The effectiveness of the collected heuristics was clear. Heuristics (e.g., *eat a hot breakfast*) are easy to comply with and more effective in making better food choices, such as suggesting *hot*-tagged items for any query with *breakfast*-related items.

The next stage is to match the description or the output of the rule to the knowledgebase entries. The knowledge base is a domain ontology consisting of classes, relationships and instances of classes. For instance the sample ontology used as an example in this paper 'Fruit' and 'Juice' are classes and between them there exists a relation like 'hasForm' and its inverse 'isFormedBy'. We define a concept as being a class or an instance of a class, such as 'Banana' is an instance of 'Fruit'.

User profile is constructed by calculating TF-IDF values for each term. We determine the term frequency (TF) $f_{i,j}$ for a term $t_i$ within an recipe $a_j$:

$$tf_{i,j} = \frac{n_{i,j}}{\sum_k n_{k,j}} \quad (1)$$

dividing $n_{i,j}$, the number of occurrences of term $t_i$ in recipe by $a_j$, the total number of terms in the document. Then the inverse document frequency (IDF):

$$idf_i = \log \frac{|A|}{|\{a : t_i \in a\}|} \quad (2)$$

dividing the total number of food items by the number of food items containing term $t_i$. The final value is computed by multiplying TF and IDF:

$$tfidf_{i,j} = tf_i \times idf_i \quad (3)$$

Semantic measures benefit from the ontology that is defined by a set of concepts:

$$C = \{c_1, c_2, c_3, \cdots, c_n\} \quad (4)$$

The food recipe can be defined by a set of $p$ concepts:

$$A = \{c_1^a, c_2^a, c_3^a, \cdots, c_p^a\} \quad (5)$$

The user profile, $U$, consists of $q$ concepts found in the food items read by the user:

$$U = \{c_1^u, c_2^u, c_3^u, \cdots, c_q^u\} \quad (6)$$

The similarity between a food recipe and the user profile can be computed by:

$$Similarity(U, A) = \begin{cases} 1 & \text{if } |U \cap A| > 0 \\ 0 & \text{otherwise} \end{cases} \quad (7)$$

We can employ binary cosine to compute the similarity:

$$B(U, A) = \frac{|U \cap A|}{|U| \times |A|} \quad (8)$$

by dividing the number of concepts in the intersection of the user profile and the unread food recipe by the product of the number of concepts in respectively *U* and *A*.

Similarly, Jaccard computes the similarity between two sets of concepts:

$$J(U, A) = \frac{|U \cap A|}{|U \cup A|} \quad (9)$$

Semantic neighborhood of $c_i$ is all concepts directly related to $c_i$ including $c_i$:

$$N(c_i) = \{c_1^i, c_2^i, c_3^i, \cdots, c_n^i\} \tag{10}$$

A food item $a_k$, which consists of $m$ concepts is described as the following set:

$$A_k = \{c_1^k, c_2^k, c_3^k, \cdots, c_m^k\} \tag{11}$$

To compare two new items $n_i$ and $n_j$, a vector can be created:

$$V_l = \left(\langle c_1^l, w_1^l \rangle, \cdots, \langle c_p^l, w_p^l \rangle\right) \qquad l \in \{i, j\} \tag{12}$$

where $w_i$ is the weight of $c_i$. The similarity between food items $a_i$ and $a_j$ is :

$$\mathrm{SemRel}(a_i, a_j) = \cos(V_i, V_j) = \frac{V_i \cdot V_j}{\|V_i\| \times \|V_j\|} \in [0,1] \tag{13}$$

The proposed framework is implemented in Java. It allows the user to formulate queries and execute them to retrieve relevant food items. We use the approach applied to adaptive hypermedia [21] and Hermes framework[22]. Hermes framework was originally used for building personalized *news* services. We extend Hermes with food recommendation functionality. It utilizes OWL[23] for representing the ontology.

Performed tests are based on a corpus of 300 food items extracted from the United States Department of Agriculture (USDA) [24] as shown in Table 1.

Table 1.food database

| Group | No. of items | Group | No. of items |
|---|---|---|---|
| American Indian | 165 | Lamb and Veal | 345 |
| Baby Foods | 329 | Legumes | 386 |
| Baked Products | 497 | Nut and Seed | 128 |
| Beef Products | 757 | Pork Products | 340 |
| Beverages | 284 | Poultry Products | 388 |
| Breakfast Cereals | 408 | Restaurant and Meals | 121 |
| Cereal Grains | 184 | Sausages and Luncheon | 234 |
| Dairy and Egg | 253 | Snacks | 169 |
| Fast Foods | 385 | Soups and Sauces | 510 |
| Fats and Oils | 220 | Spices and Herbs | 61 |
| Finfish | 258 | Sweets | 341 |
| Fruits and Juices | 329 | Vegetables | 814 |

We have used 5 users with different but well-defined interests in our experiments. An example of a user interest is "Fruits". Each user has manually rated the food items as relevant or non-relevant for his interest. For each user we split the food items corpus in two different sets: 60% of the food items are the training set and 40% of the food items are the test set. Recommenders compute the similarity between the food items and previously computed user profile. If the computed similarity value is higher than a predefined cut-off value the food item is recommended and ignored otherwise. Evaluating the recommenders is done by measuring accuracy, precision, recall, specificity, and F-measure. This is done by calculating a confusion matrix for each user. Table 2 shows the results of the evaluations and Fig. 2 visualizes them.

Table 2. Evaluation results

|  | Accuracy | Precision | Recall | Specificity | F-Measure |
|---|---|---|---|---|---|
| **TF-IDF** | 90% | 90% | 45% | 99% | 60% |
| **B. Cosine** | 47% | 23% | 95% | 36% | 37% |
| **Jaccard** | 93% | 92% | 58% | 99% | 71% |
| **Sem. Rel.** | 57% | 26% | 92% | 47% | 41% |
| **Proposed** | 94% | 93% | 62% | 99% | 74% |

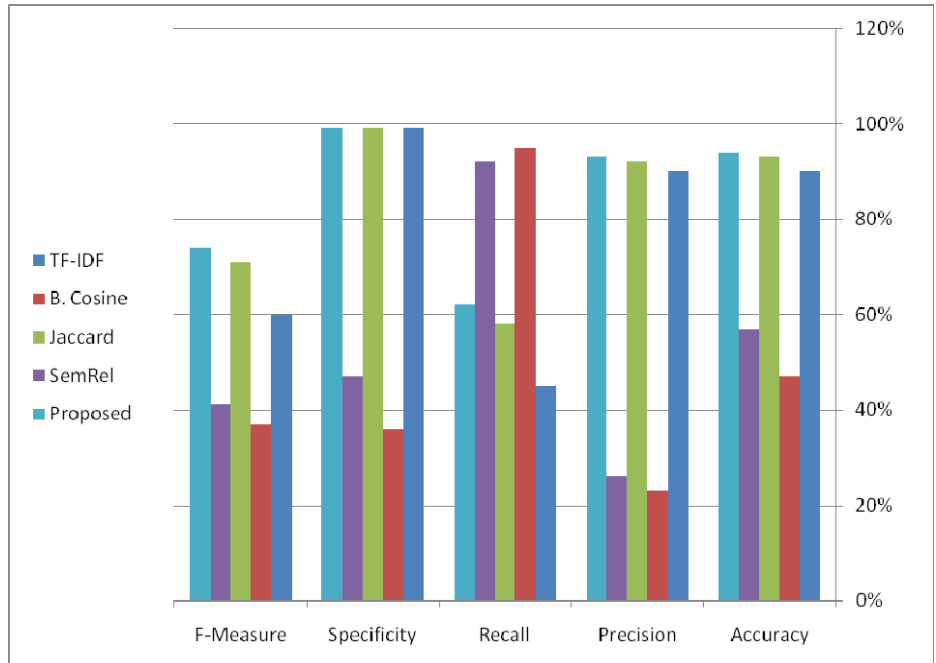

**Fig. 2**. Evaluation results

The best recommenders for accuracy is the proposed framework, for precision is the proposed framework, for recall is binary cosine, for specificity are TF-IDF, Jaccard, and the proposed framework, and for F-measure is the proposed framework. The proposed algorithm scores well on accuracy as it makes relatively small amount of errors for both recommended food as well as discarded food items. For precision, the proposed algorithm scores the best for precision as most recommended food items are relevant. The good results for recall obtained by the concept equivalence are due to the optimistic nature of the algorithm: any food item which involves previously viewed concepts is recommended. TF-IDF, Jaccard, and the proposed framework score well on specificity as these algorithms do not recommend most of the non-relevant food items.

## 4      Conclusion and future work

The framework can be used for building a personalized nutrition service. Based on a set of concepts, selected by the user, it is able to determine which items are relevant.

The knowledge base is a domain ontology consisting of classes, relationships and instances of classes. The knowledge base has initially been extracted from the United States Department   of Agriculture (USDA) provided a comprehensive food database. Based on the performed evaluation, we conclude that semantic recommender systems in general outperform traditional recommenders systems with respect to accuracy, precision, and recall, and that the proposed recommender has a better F-measure than existing semantic recommenders.

In the future we plan to extend the querying language by defining its grammar, and applying it for extracting deep knowledge from food ontology.

Another possible research direction relates to the advanced traditional weighting schemes that other than TF-IDF such as logarithmic TF functions [25]. Another research direction is the considered similarity function. We would like to evaluate alternatives for cosine similarity as Lnu.ltu [26] which seem to remove some of the cosine similarity bias favoring long documents over short documents.

As additional further work we would like to consider other types of food recommendation services as collaborative filtering or hybrid approaches. Also, we would like to investigate the performance of this type of recommenders with respect to existing hybrid recommenders.